\documentclass[12pt]{article}

\usepackage{amsmath}
\usepackage{amsfonts}
\usepackage{amssymb}
\usepackage{amsthm}
\usepackage[nocomma]{optidef}
\usepackage[T1]{fontenc}

\theoremstyle{definition} %avoid italics in remarks.
\newtheorem*{remark}{Remark}
\newtheorem{definition}{Definition}

\usepackage[numbers]{natbib}
\usepackage{hyperref}

\usepackage{graphicx}
\graphicspath{{fig/}{Fig/}{../out/}{}}

\title{Comprehensive identifiability analysis and reliable parameter estimation for an SEIR model}
\author{Eduard Campillo-Funollet, James Van Yperen} %Alphabetical.

\date{\today}

\newcommand{\bN}{\beta_N}

\begin{document}

\maketitle

\abstract{
The Susceptible-Exposed-Infectious-Removed (SEIR) model is a fundamental model
in epidemiology. Model parameters such as the reciprocal transmission,
incubation, and infectious rates are often difficult to measure directly, and
they are estimated by solving an optimisation problem aiming to minimise the
difference between the observed data and the model solution. However, the
parameters of the standard SEIR system are not globally identifiable, causing
optimisation algorithms to frequently converge to incorrect local optima and
suffer from numerical stiffness. Here we show a comprehensive structural
identifiability analysis of the SEIR framework, and present a globally
identifiable and computationally stable reparameterisation of the model derived
via an observational system approach. We fully characterise the multiple
locally identifiable parameters, and by transforming the system into a globally
identifiable structure, we eliminate the non-uniqueness issues in the parameter
estimation approaches. Our numerical experiments demonstrate that this
reformulation significantly improves convergence frequency, avoids runtime
errors caused by numerical overflow, and consistently recovers the correct
parameters. Furthermore, incorporating first-order sensitivity equations into
the optimiser enhances the robustness and execution speed of the estimation
process. Numerically well-conditioned methods for parameter identification,
together with a comprehensive understanding of the identifiability of the
parameters, ensure that the model yields reliable, rigorous insights for
infectious disease forecasting and theoretical epidemiology.

}

\section{Introduction}

Susceptible-Exposed-Infectious-Removed (SEIR) models are the cornerstone of numerous compartmental epidemiological models. Originally introduced by \citet{cooke67}, it was studied as an approach to incorporate a delay between the onset of an infection and becoming infectious \citep{burke24}. SEIR models were extensively used during the COVID-19 pandemic \citep{estrada20}, either on their own or as part of larger models \citep{wu20b}, for example with age-structure \citep{prem20}, and are regularly used to model influenza \citep{brauer19,dukic12,mikolajczyk09}. There are many alternative flavours of compartmental models, which are used as analytical tools by academics or tools for policy, see the following and references therein \citep{anastassopoulou20,hu25,rock14,siegenfeld22,sweileh22,tang20}. 

Let $S,E$, $I$ and $R$ represent respectively the number of susceptible, exposed, infectious and removed individuals. Given some initial conditions $S(0) = S_0$, $E(0) = E_0$, $I(0) = I_0$ and $R(0) = R_0$, the system of ordinary differential equations (ODEs) that characterises the evolution of $S$, $E$, $I$ and $R$ is 
\begin{align}
S'(t) &= -\bN S(t) I(t), \label{eqn:S}\\
E'(t) &=  \bN S(t) I(t) - \alpha E(t), \label{eqn:E}\\
I'(t) &= \alpha E(t) - \gamma I(t), \label{eqn:I} \\
R'(t) &= \gamma I(t). \label{eqn:R}
\end{align}
The total population size $N = S(t) + E(t) + I(t) + R(t)$ remains constant over time. With $\bN = \beta N^{-1}$, we denote $\beta$, $\alpha$ and $\gamma$ as the reciprocal of the average transmission period, average incubation period, and average infectious period respectively.

In order to compare the solution to \eqref{eqn:S}--\eqref{eqn:R} with empirical observations, we need to determine the values of the parameters $\beta_N$, $\alpha$ and $\gamma$, and the initial conditions. A common approach to this challenging task is to frame it as a minimisation problem, for example using a least squares approach aiming to minimise the difference between the observed data and the model solution \citep{aster18}. Structural identifiability characterises the existence and uniqueness of such minimisers, and for ODE systems it can be determined algorithmically in the absence of observation error, see for example \citet{saucedo24} and references therein. More precisely, to demonstrate structural identifiability, one studies the mapping from the observable quantities to the parameters of interest, usually by means of an inverse function theorem, see \citet{cunniffe24}, to determine whether the parameters are globally identifiable, locally identifiable or not identifiable. Global identifiability implies that there is a unique parameter for a given observable dataset.

The parameters in the SEIR system \eqref{eqn:S}--\eqref{eqn:R} are not globally identifiable. If an optimisation algorithm converges when fitting the model, it may converge to any of the locally identifiable parameter sets. Furthermore, the optimisation algorithm may preferentially converge to a particular local optimum, due to the complex shapes, and differing sizes, of the corresponding basins of attraction. In Section \ref{sec:experiment} we provide a complete example of a parameter identification problem for the system \eqref{eqn:S}--\eqref{eqn:R}, where initial guesses close to the parameter value that generates the data converge to a different local optimum. To address these issues, we study the identifiability problem in Section \ref{sec:analysis} to completely characterise the identifiability of the parameters, and we propose a globally identifiable reparameterisation of \eqref{eqn:S}--\eqref{eqn:R}. Interestingly, we show that some relevant quantities, such as the effective reproduction number, are globally identifiable, even in the presence of unknown under-reporting rates and the aforementioned lack of global identifiability for the SEIR model. Globally identifiable reparameterisations can be obtained algorithmically, but are not unique; in fact, we provide two in this paper. In Section \ref{sec:fituvy} we show that our proposed reparameterisation is numerically favourable, avoiding runtime errors in the ODE solver due to stiffness or numerical overflow, and that on average it converges to the optimum parameters much faster.

The identifiability and observability properties of general ODE models has been extensively studied in the literature, with some further focus based on epidemic models \citep{audoly01,bearup13,brouwer17,dankwa22,evans05,kao18,perasso11,villaverde19}. These papers demonstrate systematic ways of obtaining the identifiability properties of an ODE system, whether that be by hand or by use of computational algebra packages. Many of these techniques have subsequently been implemented in different programming languages, such as in \citet{bellu07,dong23,hong19,karlsson12,ligon17,meshkat14,sedoglavic01,villaverde16}. For a comprehensive review of identifiability methods, we refer readers to the following papers \citet{chis11,cunniffe24,hong20,raue14}.

Our identifiability analysis in this paper is based on the input-output approach proposed by \citet{bellman70}, whereby we derive a new ODE that characterises the observable quantities, in terms of the parameters and observable state. The observational system, as we call it, provides us a method to determine the identifiability properties of the ODE model. This approach has been used in susceptible-infectious-removed models (SIR), including the analysis of a boundary value problem for the observational system \citep{campillo22}, and we extensively used the approach to calibrate an SEI*R-D type model used to forecast COVID-19 hospital admissions, demand, discharges, and deaths, as well as deaths outside of hospital, during the COVID-19 pandemic \citep{campillo21}.

\section{Parameter estimation issues}\label{sec:experiment}

Consider a dataset generated by solving \eqref{eqn:S}--\eqref{eqn:I} with initial conditions and parameters listed in Table \ref{tab:exact}, and assume that from this simulation we observe $y_i = I(t_i)$, for $t_i = 0,1,\ldots,49$. Since we consider observable data on the $I$ compartment, equation \eqref{eqn:R} does not influence the observations. We note here that we have set $N=1$, which means that we are simulating the proportion of individuals who are infectious, rather than the number of infectious individuals. We aim to infer the value of the parameters $\alpha$ and $\gamma$ from the observables $y_i$, as well as the initial conditions $S_0$ and $E_0$, using non-linear least squares, namely
\begin{mini}|l|[3]
{S_0,E_0,\alpha,\gamma}{\sum_{i=1}^{49} (y_i - I(t_i))^2 \qquad}
{}{}
\addConstraint{\left\{\begin{array}{lc}
    S'(t) = -0.2 S(t) I(t), & S(0) = S_0, \\
    E'(t) = 0.2 S(t) I(t) - \alpha E(t), & E(0) = E_0, \\
    I'(t) = \alpha E(t) - \gamma I(t), & I(0) = 0.1,
\end{array} \right.}
\addConstraint{S_0,E_0,\alpha,\gamma\geq0.}
\label{eqn:ls}
\end{mini}
In this example, and to simplify the presentation, we will fix $\beta_N$ and $I_0$ in the minimisation algorithm since they are both globally identifiable. In later sections we consider the problem of identifying all parameters and initial conditions at once.

\begin{table*}[hbt!]
\caption{Initial conditions and parameter values used to generate the data for the example in Section \ref{sec:experiment}.} \label{tab:exact}
\centering
\begin{tabular}{|| c || c | c | c | c | c | c ||} 
\hline
Variable & $S_0$ & $E_0$ & $I_0$ & $\beta_N$ & $\alpha$ & $\gamma$ \\
\hline
Value &  0.7  &  0.2  &  0.1  &  0.2      &  0.3     &  0.1 \\  
\hline
\end{tabular}
\end{table*}

To obtain the results, we solve the system of ODEs using the LSODA solver (with default parameters) implemented in the function \texttt{odeint} provided by the Python package \texttt{scipy.integrate}, and we solve the minimisation problem by using the Nelder-Mead algorithm (with default parameters) implemented in the function \texttt{minimize} provided by the Python package \texttt{scipy.optimize} \citep{virtanen20}. We selected these methods and algorithms as examples of commonly used ODE solvers and optimisation algorithms. We use the numerical structures provided by \texttt{numpy} \citep{harris20}, \texttt{pandas} \citep{mckinney10} to organise the results in dataframes, and \texttt{matplotlib} \citep{hunter07} for visualisation.

\begin{figure}[hbt!]
\includegraphics[width=\linewidth]{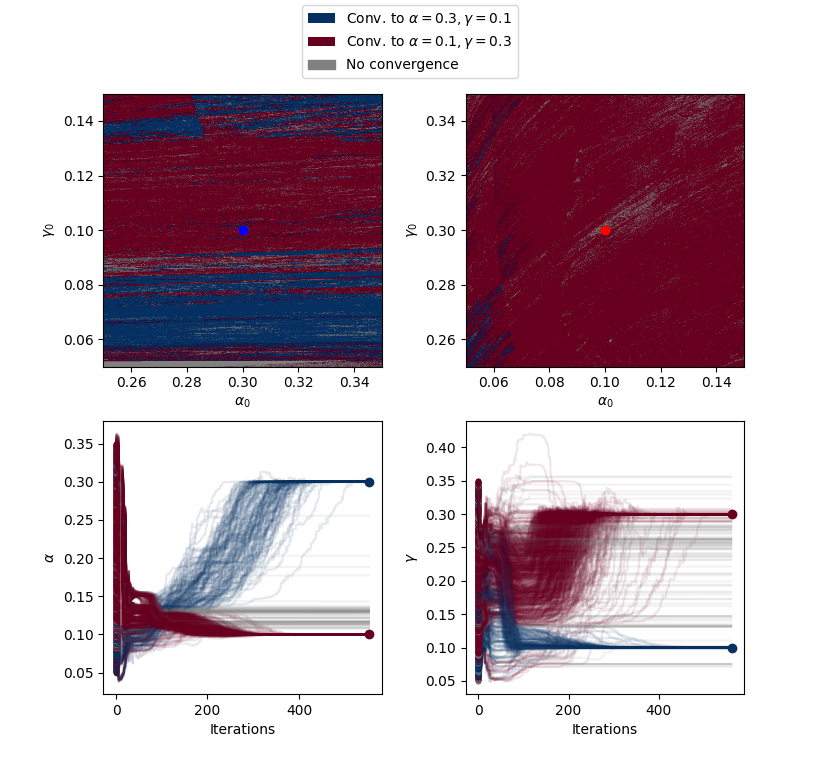}
\caption{Results of the minimisation problem \eqref{eqn:ls}. Blue denotes convergence to the minimiser $(\alpha,\gamma) = (0.3, 0.1)$, red denotes convergence to the minimiser $(\alpha,\gamma) = (0.1, 0.3)$ and grey denotes a parameter set whereby the algorithm did not converge. Top row: close up of the basins of attraction of the minimisers $(\alpha,\gamma) = (0.3, 0.1)$ (left) and $(\alpha,\gamma) = (0.1, 0.3)$ (right), where $(\alpha_0,\gamma_0)$ denote the initial guesses to the minimisation algorithm. Bottom row: traces of $\alpha$ (left) and $\gamma$ (right) as the minimisation algorithm progresses.} 
\label{fig:fit_ag_convergence}
\end{figure}

To explore the convergence of \eqref{eqn:ls}, we generate $10^6$ initial guesses for the minimisation algorithm and track the values of $\alpha$, $\gamma$, $S_0$ and $E_0$ as the algorithm converges towards a minimiser. For ease of demonstration, we fix the initial guesses for $S_0$ and $E_0$ to $1.4$ and $0.5$ respectively, and we consider a uniform grid on $[0.05,0.35]^2$ of size $10^3\times10^3$ for $\alpha$ and $\gamma$. The results of the minimisation problem \eqref{eqn:ls} are presented in Figures \ref{fig:fit_ag_convergence}. We remark that we characterised convergence by setting a threshold for the least squares minimum of \eqref{eqn:ls}: we consider that an initial guess converged if the output of the minimisation algorithm gives a value for the objective function less or equal to $10^{-6}$. We observe three different behaviours in Figure \ref{fig:fit_ag_convergence} depending on the initial guess: convergence to $(\alpha,\gamma) = (0.3, 0.1)$, convergence to $(\alpha,\gamma) = (0.1, 0.3)$, or non-convergence, mostly due to the minimisation algorithm stopping because of slow progress. 

From the top panels in Figure \ref{fig:fit_ag_convergence}, we see that initial guesses for the minimisation algorithm corresponding to small perturbations from the true solution $(\alpha,\gamma) = (0.3,0.1)$ often result in convergence to the wrong parameter set $(\alpha,\gamma)=(0.1,0.3)$. Furthermore, the second parameter set seems to be much more attractive: about 80\% of the initial guesses result in a convergence to the second, and wrong, parameter set.

In order to address the convergence issues caused by the presence of two local minimisers of the objective function, we will study in detail the identifiability of the parameters and initial conditions in \eqref{eqn:S}--\eqref{eqn:I}. 

\section{Identifiability analysis} \label{sec:analysis}

The objective of this section is to study the identifiability and observability properties of the SEIR model, and to derive a reparametrised model that is globally identifiable. The method presented here, which we call the observational system, relies on algebraic manipulation of the ODE system to obtain a new ODE with the observable data as the dependent variable. We will derive the observational system for the SEIR model, demonstrate how one can use it to determine structural identifiability properties of the parameters, and how it provides insight to deciding how to reparametrise the SEIR model.

As in Section \ref{sec:experiment}, we define the observable as
\begin{align}
    y(t) := I(t). \label{eqn:y}
\end{align}
The results in this section can also be derived including an unobserved parameter, say $k$, as a coefficient to $I$ in \eqref{eqn:y}, that can model for example an unknown under-reporting rate $k = r$, or we could set $k = \gamma$ to model situations where individuals stop being infectious once they are observed, for example when quarantine measures are in place \citep{capaldi12,hadeler11,hsieh10,pollicott11,roda20}. For ease of exposition, we are not including $k$ in the derivation, but we will discuss how the results change with the inclusion of $k$. There are many other types of observable used in the literature, such as cumulative infections \citep{cunniffe24,liu20,magal18},
\begin{align*}
    y(t) = \gamma \int_{0}^t I(s) \, \text{d}s,
\end{align*}
or the incidence \citep{dankwa22,evans05,saucedo24,tuncer18},
\begin{align*}
    y(t) = \bN S(t) I(t).
\end{align*}
The results in this paper cover the identifiability of the parameters in the case of cumulative infections, but the identifiability of the parameters from the incidence using our method is future work. 

\subsection{Observational system for SEIR}

To derive the observational system, we will obtain equations for $S$ and $E$ purely in terms of $y$, and then combine them to obtain a higher-order ODE involving only $y$ and the parameters. The coefficients of the terms in this ODE will be functions of the parameters, and these coefficients are globally identifiable \citep{cunniffe24}. Using the functions of parameters that are globally identifiable, we scale each of the compartments to also make them identifiable, thus obtaining the globally identifiable reformulation of the SEIR model. For this derivation, we do not explicitly write the dependence on $t$ for the compartments, e.g. $S = S(t)$.

By taking the first derivative of \eqref{eqn:y} and using \eqref{eqn:I}, we see that
\begin{align}
    \label{eqn:dy}
    y' = I' = \alpha \, E - \gamma \, I = \alpha \, E - \gamma \, y,
\end{align}
and hence
\begin{align}
    E = \frac{y' + \gamma \, y}{\alpha}. \label{eqn:Ey}
\end{align}
By taking the second derivative of \eqref{eqn:y} and using \eqref{eqn:E}, \eqref{eqn:I} and \eqref{eqn:Ey} we have
\begin{align}
    \label{eqn:ddy}
    y'' &= I'' = \alpha \, E' - \gamma \, I' = \alpha \left( \bN \, I \, S - \alpha \, E \right) - \gamma y' \nonumber \\
        &= \alpha \, \bN \, y \, S - \alpha(y' + \gamma \, y) - \gamma y' = \alpha \, \bN \, y \, S - (\alpha + \gamma) y' - \alpha \, \gamma \, y,
\end{align}
and hence
\begin{align}
    S = \frac{y'' + (\alpha + \gamma) y' + \alpha \, \gamma \, y}{\alpha \, \bN \, y}. \label{eqn:Sy}
\end{align}
By adding \eqref{eqn:S}--\eqref{eqn:I} we have
\begin{align}
    S' + E' + I' = -\gamma \, I, \label{eqn:conservation}
\end{align}
which, when combined with the first derivatives of \eqref{eqn:Ey} and \eqref{eqn:Sy}, gives
\begin{align*}
     - \gamma \, y  &= \frac{1}{\alpha \, \bN} \left(\frac{y''}{y} \right)' + \frac{\alpha + \gamma}{\alpha \, \bN} \left( \frac{y'}{y} \right)' + \frac{1}{\alpha} y'' + \frac{\gamma}{\alpha} y' + y'.
\end{align*}
Multiplying by $\alpha \, \bN $ leaves the observational model
\begin{align}
    \label{eqn:obs}
    \left(\frac{y''}{y} \right)' = - (\alpha + \gamma) \left( \frac{y'}{y} \right)' - \bN y'' - \bN(\alpha + \gamma) y' - \bN \, \alpha \, \gamma \, y.
\end{align}

\begin{remark}
    If one was to consider the observables $y(t) = k I(t)$, then the observational model becomes
    \begin{align*}
        \left(\frac{y''}{y} \right)' = - (\alpha + \gamma) \left( \frac{y'}{y} \right)' - \frac{\bN}{k} y'' - \frac{\bN}{k}(\alpha + \gamma) y' - \frac{\bN}{k} \, \alpha \, \gamma \, y.
    \end{align*} 
\end{remark}

\subsection{Parameter identifiability using the observational system}

Before concluding about the identifiability properties, let us provide a definition of identifiability, which we have adapted from \citet{linayage26}. 

\begin{definition}[Structural Identifiability]
    Let $p$ and $\hat{p}$ be distinct model parameters (including initial conditions), and let $y(t,p)$ be the observations. If
    \begin{align}
        \label{identifiability}
        y(t,p) = y(t,\hat{p}) \implies p = \hat{p}
    \end{align}
    then we conclude that the model is \emph{structurally identifiable} from noise-free and continuous observations $y(t)$. If \eqref{identifiability} is true globally, then we call the parameters \emph{globally identifiable}. If \eqref{identifiability} is only true for a $\hat{p}$ in a neighbourhood of $p$ then we call the parameters \emph{locally identifiable}.
\end{definition}

As aforementioned in the previous section, the coefficients of the terms of the observational model are globally identifiable. From the coefficients of \eqref{eqn:obs}, we can expect to identify the parameters 
\begin{align}
    \label{eqn:c}
    c_1 := \alpha + \gamma, \qquad c_2 := \bN, \qquad c_3 := \alpha \, \gamma.
\end{align}
$I_0$ is globally identifiable due to the description of the data, but $S_0$ and $E_0$ are only locally identifiable. We see this by inspecting the equations \eqref{eqn:Ey} and \eqref{eqn:Sy} and noticing that we can not write them in terms of $c_1$, $c_2$ and $c_3$. From \eqref{eqn:c} we can confirm that $\bN$ is globally identifiable, but $\alpha$ and $\gamma$ are not, since $c_1$ and $c_3$ are invariant with respect to the ordering of $\alpha$ and $\gamma$. That is, there are two parameter sets, $(\alpha,\gamma)$ and $(\gamma,\alpha)$ that produce the same values $(c_1,c_3)$, and consequently the same observable $y$. To find the initial conditions associated to the second parameter set $(\bN,\gamma,\alpha)$, let us denote $\bar{S}$ and $\bar{E}$ as the corresponding solutions of \eqref{eqn:S}--\eqref{eqn:I}. From \eqref{eqn:dy}, we have
\begin{align*}
    \alpha \, E - \gamma \, y = y' = \gamma \, \bar{E} - \alpha \, y, 
\end{align*}
which rearranges to give
\begin{align*}
    \bar{E} = \frac{\alpha}{\gamma} E + \left(\frac{\alpha}{\gamma} - 1\right) I. 
\end{align*}
Considering \eqref{eqn:ddy}, we have
\begin{align*}
    \alpha \, \bN \, y \, S - (\alpha + \gamma) y' - \alpha \, \gamma \, y = y'' = \gamma \, \bN \, y \, \bar{S} - (\gamma + \alpha) y' - \gamma \, \alpha y
\end{align*}
which reduces down to
\begin{align*}
    \bar{S} = \frac{\alpha}{\gamma} S.
\end{align*}
Hence, the following two parameter and initial condition sets will result in the exact same data
    \begin{subequations}
        \begin{align}
            \left(\bN, \alpha, \, \gamma \right) \qquad \text{ with } \qquad (S_0, \, E_0, \, I_0), \label{eqn:pset1}
        \end{align}
        and
        \begin{align}
            \left(\bN, \, \gamma, \, \alpha \right) \qquad \text{ with } \qquad \left(\frac{\alpha}{\gamma}S_0, \, \frac{\alpha}{\gamma}E_0 + \left[\frac{\alpha}{\gamma} - 1\right] I_0, \, I_0 \right). \label{eqn:pset2}
        \end{align}
    \end{subequations}

\begin{remark}
    Let us consider the case $y(t) = k I(t)$. From the previous remark, we see that only $c_2$ changes and becomes
    \begin{align*}
        c_2 = \frac{\bN}{k}.
    \end{align*}
    Let us now consider some different choices of $k$. 
    \begin{enumerate}
        \item If $k=r$, where $r$ is some unknown under-reporting rate, we can scale both $\bN$ and $k$ by some positive constant, say $\ell$, that leaves $c_2$ unchanged. This clearly shows that $\bN$ and $k$ are not identifiable. Moreover, \eqref{eqn:pset1} becomes
        \begin{align*}
            \left(\ell \bN, \, \alpha, \, \gamma, \, \ell k \right) \qquad \text{ with } \qquad \left(\frac{1}{\ell}S_0, \, \frac{1}{\ell} E_0, \, \frac{1}{\ell} I_0 \right),
        \end{align*}        
        where we can see that all the initial conditions are also not identifiable, but $\alpha$ and $\gamma$ remain locally identifiable. Lacking identifiability (even locally) is not uncommon when under-reporting is included, see \citet{cunniffe24} for example.
        \item If $k=\gamma$, then \eqref{eqn:pset2} becomes
        \begin{align*}
            \left(\frac{\alpha}{\gamma} \bN, \, \gamma, \, \alpha \right) \qquad \text{ with } \qquad \left(S_0, \, \frac{\alpha}{\gamma}E_0 + \left[\frac{\alpha}{\gamma} - 1\right] I_0, \, I_0 \right).
        \end{align*}
        Now $\bN$ is only locally identifiable, but $S_0$ becomes globally identifiable.
    \end{enumerate}
\end{remark}

\begin{remark}
    There is some flexibility in how you choose the parameter set \eqref{eqn:c}. For example, we could instead have chosen the set
    \begin{align*}
    \tilde{c}_1 := \alpha + \gamma, \qquad \tilde{c}_2 := \bN(\alpha + \gamma), \qquad \tilde{c}_3 := \bN \, \alpha \, \gamma.
    \end{align*}
    There is a question here concerning the ``best'' choice of the parameter set, for example does the optimiser behave better when estimating the parameter set given by \eqref{eqn:c} instead of what is presented here? We leave this question for further study. 
\end{remark}

\subsection{Globally identifiable reformulation of SEIR}

Using the derivation \eqref{eqn:obs}, we demonstrated how we can write the compartments in terms of the data and parameters, namely \eqref{eqn:Ey} and \eqref{eqn:Sy}. In order to obtain a globally identifiable reformulation of the SEIR model, we need to write the states (or a transformation of them) in terms of the globally identifiable parameters \eqref{eqn:c}. Namely, by adding \eqref{eqn:y} and \eqref{eqn:Ey} and scaling by $\gamma$ we have
\begin{align}
    \label{eqn:EIyc}
    \gamma^{-1} (E(t) + I(t)) = \frac{y'(t) + c_1 \, y(t)}{c_3}.
\end{align}
As for \eqref{eqn:Sy}, we only need to scale by $\gamma$ to obtain
\begin{align}
    \label{eqn:Syc}
    \gamma^{-1} \, S(t) = \frac{y''(t) + c_1 \, y'(t) + c_3 \, y(t)}{c_2 \, c_3 \, y(t)}.
\end{align}
An interesting consequence of this is
\begin{align}
    \label{ycRt}
    \mathcal{R}_t := \frac{\bN}{\gamma} S(t) = \frac{y''(t) + c_1 \, y'(t) + c_3 \, y(t)}{c_3 \, y(t)},
\end{align}
which means that we can still determine the effective reproduction number associated to the SEIR without being able to fully determine the number of susceptible individuals. We note though that the basic reproduction number, which in this case is $\mathcal{R}_0 := \bN / \gamma$, can not be observed. Relations \eqref{eqn:Syc} and \eqref{eqn:EIyc} motivate the following change of variables: setting $u = \gamma^{-1} S$, $v = \gamma^{-1} (E + I)$, and keeping $y = I$, we can manipulate \eqref{eqn:S}--\eqref{eqn:I} to become
\begin{align}
    u'(t) &= - c_2 \, u(t) \, y(t), \quad & \, u(0) = u_0, \label{eqn:uc} \\
    v'(t) &= c_2 \, u(t) \, y(t) - y(t), & \, v(0) = v_0, \label{eqn:vc} \\
    y'(t) &= c_3 \, v(t) - c_1 \, y(t) , \quad & y(0) = y_0. \label{eqn:yc}
\end{align}
We will refer to this as the UVY model. To ensure that the solution to \eqref{eqn:uc}--\eqref{eqn:yc} correspond to be a feasible set of solutions to \eqref{eqn:S}--\eqref{eqn:I}, we must require that
\begin{align}
    \label{discriminant}
    c_1^2 - 4c_3 \geq 0.
\end{align}
We note that feasible here means that there exists some real transformation between the solutions. This condition comes from the mapping from $(c_1,c_3)$ back to $(\alpha,\gamma)$, namely
\begin{align} \label{eqn:xpm}
    x_{\pm} = \frac{1}{2}\left(c_1 \pm \sqrt{c_1^2 - 4 \, c_3} \right)
\end{align}
where $x_{\pm}$ can represent either $\alpha$ or $\gamma$. If the condition is not satisfied, then the solution to \eqref{eqn:uc}--\eqref{eqn:yc} corresponds to a complex solution to \eqref{eqn:S}--\eqref{eqn:I}. 

\begin{remark}
    The reformulation presented in equations \eqref{eqn:uc}--\eqref{eqn:yc} remains unchanged when considering data of the form $y(t) = k \, I(t)$, where we would set $u = k \gamma^{-1} S$, $v = k \gamma^{-1}(E + I)$ and $y = k I$. In this case, since $c_2 = \bN/k$, no extra $k$ is added to the equations. 
\end{remark}

\begin{remark}
    The reformulation will be different if a different set of globally identifiable parameters is chosen. Furthermore, even using the same identifiable parameters, we can obtain different reformulations. For example, noticing that
    \begin{align*}
        \alpha (E + I) = y' + \gamma y + \alpha y = y' + c_1 y
    \end{align*}
    motivates us to take $v = \alpha(E + I)$ to obtain
    \begin{align*}
        u'(t) &= - c_2 \, u(t) \, y(t), \quad & \, u(0) = u_0,  \\
        v'(t) &= c_2 \, c_3 \, u(t) \, y(t) - c_3 y(t), & \, v(0) = v_0, \\
        y'(t) &= v(t) - c_1 y(t) , \quad & y(0) = y_0. 
    \end{align*}
    Like in a previous remark, there is a question here about ``best'' choice of mapping from SEIR to globally identifiable states, considering aspects such as stiffness of the ODE, which we leave to further study.
\end{remark}

\section{Fitting the reformulation of the SEIR model}
\label{sec:fituvy}

\subsection{Contrast to the previous experiment}
\label{sec:uvy_fit}

We can now repeat the experiment in Section \ref{sec:experiment} except using the reformulated version of the SEIR equations, namely 
\begin{mini}|l|[3]
{u_0,v_0,c_1,c_3}{\sum_{i=1}^{49} (y_i - y(t_i))^2 \qquad}
{}{}
\addConstraint{\left\{\begin{array}{lc}
    u'(t) = -0.2 u(t) y(t), & u(0) = u_0 \\
    v'(t) = 0.2 u(t) y(t) - y(t), & v(0) = v_0 \\
    y'(t) = c_3 v(t) - c_1 y(t), & y(0) = 0.1
\end{array} \right.}
\addConstraint{c_1^2 - 4c_3 \geq 0}
\addConstraint{u_0,v_0,c_1,c_3\geq0.}
\label{eqn:cls}
\end{mini}
To facilitate the comparison with the results in Section \ref{sec:experiment}, we fix the value of $c_2$ and $y_0$, which is equivalent to fixing $\bN$ and $I_0$ in the original formulation. The associated reformulated parameters are listed in Table \ref{tab:exact_uvy}. 

\begin{table*}[hbt!]
\caption{Initial conditions and parameter values used to generate the data for the example in Section \ref{sec:fituvy}.} \label{tab:exact_uvy}
\centering
\begin{tabular}{|| c || c | c | c | c | c | c ||} 
\hline
Variable & $u_0 $ & $v_0$ & $y_0$ & $c_1$ & $c_2$ & $c_3$ \\
\hline
Value &  7     &  3    &  0.1   & 0.4  &  0.2  &  0.03 \\
\hline
\end{tabular}
\end{table*}

\begin{figure}[hbt!]
\includegraphics[width=\linewidth]{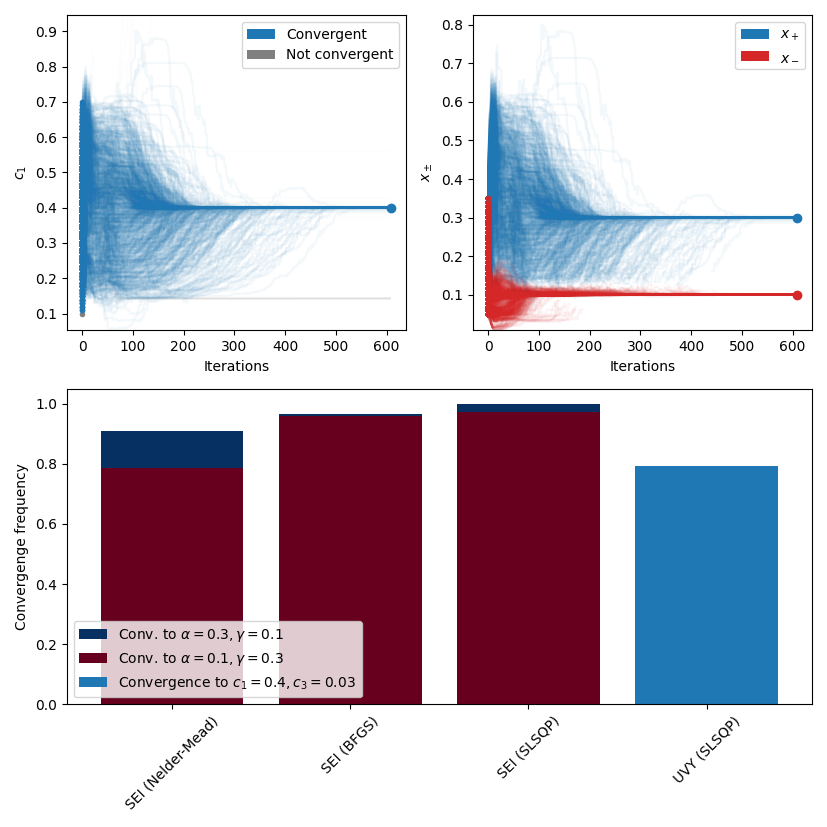}
\caption{Results of minimisation problem \eqref{eqn:cls}. Top row left: minimiser iterations for $c_1$ when fitting \eqref{eqn:uc}--\eqref{eqn:yc}. Top row right: corresponding values for $x_\pm$. Bottom row: convergence frequency of the two models and different choices of numerical methods in the optimiser.} \label{fig:fit_sei_vs_uvy_convergence}
\end{figure}

Again, like in Section \ref{sec:experiment} we characterise convergence for an initial guess by obtaining an output of the minimisation algorithm that is less than or equal to $10^{-6}$. The top left panel of Figure \ref{fig:fit_sei_vs_uvy_convergence} shows the minimiser iterations for the parameter $c_1$. In contrast to the experiment in Section \ref{sec:experiment}, we can see that now all the initial guesses converge to the same value. The top right panel in Figure \ref{fig:fit_sei_vs_uvy_convergence} shows the convergence paths for $x_+$ and $x_-$, whereby we see that we recover the original values of $(\alpha,\gamma)=(0.3,0.1)$ using \eqref{eqn:xpm}, although in this setting we only known them as an unordered pair, that is we do not know if $x_+$ corresponds to $\alpha$ or $\gamma$. The bottom panel of Figure \ref{fig:fit_sei_vs_uvy_convergence} explores the performance of different optimisation methods, and we can see that the SEIR model does converge more frequently overall, it almost certainly converges to the wrong parameter set, which is a problem the UVY model does not suffer from.

We solve the SEIR model as in Section \ref{sec:experiment}. For the UVY model, we use the same ODE solver, but we use the the Sequential Least Squares Quadratic Programming (SLSQP) algorithm (with default parameters) implemented in the \texttt{minimize} function provided by \texttt{scipy.optimize}. We use SLSQP as it can handle constraints and bounds, whilst Nelder-Mead can only handle bounds.

\subsection{Fitting all six parameters}
\label{sec:six_params}

\begin{figure}[hbt!]
\includegraphics[width=\linewidth]{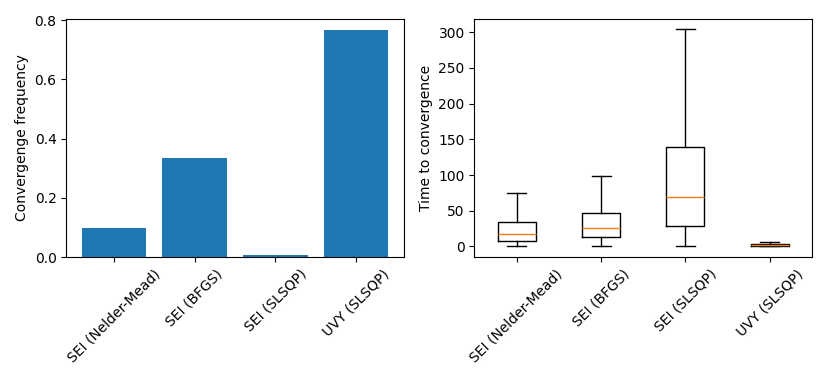}
\caption{Results of the minimisation problems \eqref{eqn:ls} and \eqref{eqn:cls}, except not fixing $\bN$ and $I_0$, and $c_2$ and $y_0$. Left: convergence frequency using different optimisation methods for the two minimisation problems. Right: comparison of time to convergence using different optimisation methods.}
\label{fig:six_params}
\end{figure}

So far we have only considered fitting four of the parameters and fixing two at the exact value to enable easier demonstration and depiction of the challenges at play. But in reality we often can not do this. Using the same parameters, and corresponding reformulated parameters, as presented in Tables \ref{tab:exact} and \ref{tab:exact_uvy} respectively, we conduct another simulation whereby we look to fit all six parameters, either $\alpha$, $\bN$, $\gamma$ and initial conditions, or $c_1$, $c_2$, $c_3$ and initial conditions. We remark that although $I_0$ is directly observable in this case, in the general setting where $y = k I$, we would need to fit that initial condition as well. The left panel in Figure \ref{fig:six_params} depicts the proportion of initial guesses that converge to a parameter set, and the right panel depicts the run time to obtain convergence. The initial guesses for the minimisation method is uniformly sampled with given lower and upper bounds for each parameter: $\bN\in[0.5,1]$, $\alpha\in[0,0.5]$, and $\gamma,S_0,E_0,I_0\in[0,1]$. The initial guesses for the UVY model are uniformly sampled from corresponding transformed region. Run time is calculated by generating initial guesses until convergence occurs. We sampled $10^6$ initial guesses for each model. An initial guess may not converge due to a run-time error, such as an overflow in the numerical integrator which typically happens in a region of stiffness, or that the optimiser reaches a maximum iteration (which we set to 2000). If an initial guess fails to converge, we resample an initial guess but continue the timer - this gives us a realistic timing for practically how long it might take to obtain a converged parameter set (but not necessarily the correct one). In contrast to the previous example, the UVY reparametrisation is significantly more reliable, in that it converges in just under 80\% of the simulations, whilst the SEIR model struggles to converge at all regardless of the numerical algorithm you choose. Moreover, the UVY model is consistently quicker to run. This can be seen as a consequence of the fact that it is more likely to converge given an initial guess, but we wanted to demonstrate that it is still quicker than running the SEIR model many times. In summary, when fitting all the parameters, the UVY model converges more reliably and faster than fitting the SEIR model. 

\subsection{Improving the numerical optimiser}
\label{sec:optimisers}

There are many options and tolerances in the numerical methods presented that can be tuned to improve convergence frequencies, but by doing this you have to weigh against time to convergence as it typically means more iterations of your numerical method. An alternative method is to reduce the numerical errors in the method by calculating the analytical derivative of the objective function via the use of the first-order sensitivity equations rather than approximating the derivative using finite differences. The first order sensitivity equations are calculated via the ODEs characterising $\partial_{u_0} u$, $\partial_{v_0} u$, $\partial_{y_0} u$, $\partial_{c_1} u$, $\partial_{c_2} u$, and $\partial_{c_3} u$, and similarly for the partial derivatives of $v$ and $y$. Overall, the system \eqref{eqn:uc}--\eqref{eqn:yc}, together with the first-order sensitivity equations, form a system of 21 equations: three original equations, plus three equations for each one of the partial derivatives with respect to the three initial conditions $u_0$, $v_0$, and $y_0$, and the three parameters $c_1$, $c_2$, and $c_3$. The first-order sensitivity equations can be calculated by hand, which we do in the code, or one can use a library, such as \texttt{CVODES} in \texttt{SUNDIALS} or \texttt{ForwardDiff} in \texttt{Julia} \citep{gardner22,revels16}. 

\begin{figure}[hbt!]
\includegraphics[width=\linewidth]{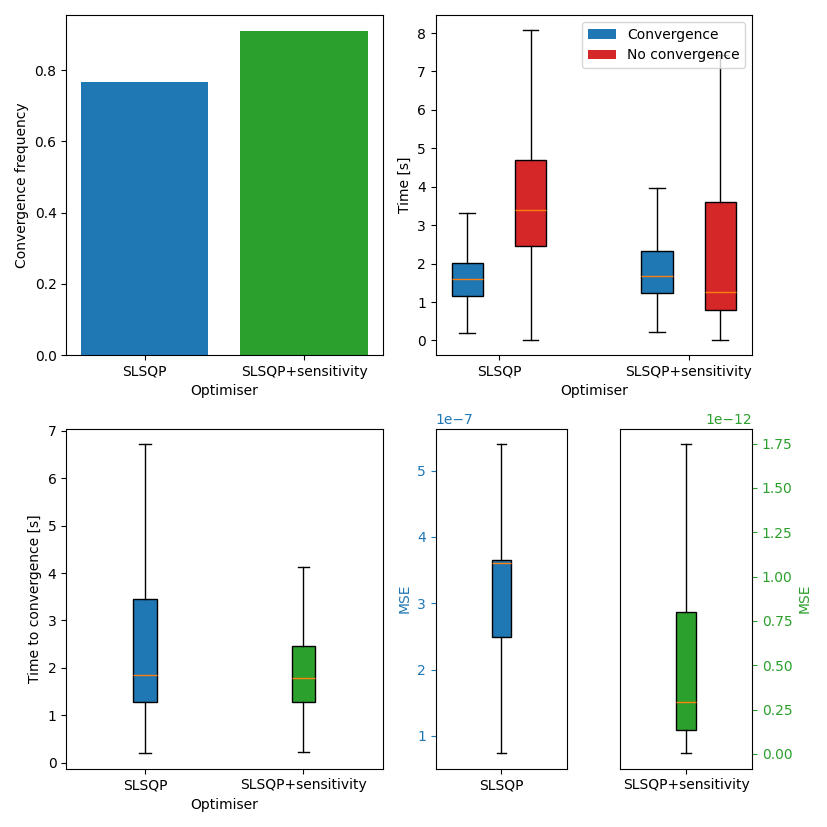}
\caption{Results of the minimisation problems \eqref{eqn:cls}, including the fitting of $c_2$ and $y_0$, and comparing the performance with and without the first-order sensitivity equations. Top left: convergence frequency. Top right: execution time in seconds for both algorithms, for initial guesses that converge, and initial guesses that fail to converge. Bottom right: total time to convergence. Bottom right: mean squared error of the estimates with respect to the true underlying parameters and initial conditions from Table \ref{tab:exact_uvy}.} \label{fig:time}
\end{figure}

Using the corresponding reformulated parameters presented in Table \ref{tab:exact} for the UVY model, we conduct another simulation whereby we include the first-order sensitivity equations. The results are presented in Figure \ref{fig:time}, where we are comparing the results using just SLSQP and using SLSQP with the first-order sensitivity equations. The top left panel demonstrates that including the first-order sensitivity equations improves the convergence frequency of the UVY model to approximately $90\%$. The bottom left panel demonstrates that the average time to convergence is roughly the same, but variation in convergence time is smaller. The top right panel demonstrates the runtime for one initial guess split by whether the initial guess converges or not. We see that the runtime when an initial guess converges is roughly the same between the two methods, but adding the first-order sensitivity equations dramatically improves the time when an initial guess does not converge. This is likely because the first-order sensitivity equations are more sensitive to areas in the parameter space that cause the numerical integrator to crash. The bottom right panel depicts the mean-squared error of the numerical solution to the optimisation problem and the true solution, where clearly the addition of the first-order sensitivity equations obtains a better fit. The addition of the first-order sensitivity equations to the optimiser provides a more reliable and more accurate numerical solver. 

\clearpage

\section{Discussion}
\label{sec:discussion}

There are well-established general results about the structural identifiability of parameters and initial conditions of ordinary differential equations. However, in the practical application of ODE models we can get a more nuanced understanding of the parameter estimation problem by analysing the specific problem at hand. Here, we studied the SEIR model, a widely applied model in epidemiology with parameters that are notoriously difficult to measure experimentally. The SEIR model is considered to be one of the simplest models one could take, and thus not necessarily representative of reality, and yet even in the simplest cases one needs to be careful with conducting their parameter estimation. The issues we identified, such as non-uniqueness and lack of solver robustness, will get exacerbated when using more complex epidemiological models. 

The numerical experiment summarised in Figure \ref{fig:fit_ag_convergence} highlights a critical issue: for the vast majority of initial guesses, we are getting convergence to a local minimum of the objective function corresponding to the wrong parameter set. We remark that even if we try a few different initial guesses, we may not even see convergence to the correct local minimum. For example, in the example presented in Section \ref{sec:experiment}, 10 random guesses will only converge to the right local minimum about 30\% of the time. Therefore, a direct approach where one fits \eqref{eqn:S}--\eqref{eqn:I} with random initial guesses may suggest that there is only one local minimum! Furthermore, in regimes with contrasting values of $\alpha$ and $\gamma$, this could lead to inaccurate conclusions on the average lengths of the exposed and infectious stages of the disease. The experiment also highlights that other common approaches for overcoming the lack of global identifiability may not work. One may like to constrain the feasible parameter space to remove one of the parameter sets (e.g., in this case one could constrain the problem with $\gamma > \alpha$), however the minimisation algorithm will likely converge to the active set of the constraint even with the correct parameter set within the feasible space. The optimisation algorithm would then suggest that, for example, $\alpha=\gamma$ or $E_0 = 0$. Alternatively, one could randomly sample the parameter space to see what the algorithms converge to, but without any prior information on the parameters one has to make a decision on which parameter set to choose. 

The identifiable reparametrisation \eqref{eqn:uc}--\eqref{eqn:yc} addresses these issues, and gives us a complete understanding of the information about parameters and initial conditions that are identifiable from the observable data. It also provides us with practical improvements, as demonstrated in Figures \ref{fig:fit_sei_vs_uvy_convergence} and \ref{fig:six_params}. One could argue that we could improve the robustness and reliability of the SEIR approach using the first-order sensitivity equations, like we do in Section \ref{sec:optimisers}, however this would not solve the main issue of the lack of global identifiability in the parameters. 

Although the implementation is specific to the SEIR model, the two key steps in our approach are general: find a globally identifiable parametrisation of the dynamical model, and exploit the sensitivity equations for numerical stability. We highlight again that there are now computational packages that can help diagnose and identify issues with parameter identifiability and help reformulate the equations (e.g., \citet{dong23}), however we note what is provided is only one set of parameters and reformulated states where there are likely many such sets, some of which could be numerically or statistically advantageous. Similarly, there are packages that can solve the first-order sensitivity equations without the need to directly implement them, see for example \citet{gardner22}. 

The method and analysis we presented here can be conducted on most systems of ODEs, particularly those that arise in biology and are in the form of a chemical reaction network. However, the more complex the original system is, the more complex the derivation can get. One can turn to computational tools to conduct algebraic manipulations to obtain the observational system, however these can also start to struggle if the system gets large. Similarly, there are alternative methods of checking for parameter identifiability that are well suited for computational algebra packages, such as utilising the Taylor's expansion method \citep{bearup13,pohjanpalo78}.

The experiment in Section \ref{sec:experiment} was conducted on a parallel setting, using 120 processing threads, taking approximately 12 hours to run. We only need this significant computational power to run many trials to demonstrate clearly the challenges of the SEIR model and improvements of the UVY model; the fitting could be conducted on a laptop in a short space of time, and in particular when using the UVY system with sensitivity equations, the computational needs of the method are very low. 

The globally identifiable parameters, as well as the globally identifiable reparameterisation of the ODE model, are not unique. A question to explore for the future is how can we determine the optimal choice of parameters and reparameterisation. There are several measures we could try to optimise. From a parameter identification perspective, we would like to minimise the error in the estimated parameters, which would correspond to optimising the sensitivity of each parameter to errors in the observables. From a statistical point of view, we would at estimates for the variance of the fitted parameters, such as by using the Fisher Information Matrix, and choose the reparameterisation that minimises it. Finally, from a numerical perspective, we would favour reparameterisations that yield well-behaved systems of ODEs, to avoid problems that cause numerical issues when exploring the parameter space due to, for example, stiffness. An interesting open question for future work is how to balance this different measures, and in particular, what is the interplay between minimising the error or variance, and maximising the numerical stability of the ODEs.

We restricted this study to structural issues, in particular to the case of exact observations. Further issues of statistical nature arise when fitting noisy observations. Furthermore, structural identifiability relies on assumption that at a fixed point in time, the time derivatives of the observable quantities are also available. However, in practical applications in epidemiology, the data is only available at discrete time points. The statistical issues, as well as the impact of the number of observations on the identifiability of the parameters, will be the focus of future studies.

\clearpage

\bibliography{ref}

\end{document}